\documentclass[aps,prd,superscriptaddress,amsmath,twocolumn,10pt,nofootinbib]{revtex4}
\usepackage{color,hangcaption,hhline,psfrag,rotating,amssymb}
\usepackage[hang,nooneline]{subfigure}
\usepackage{dcolumn}
\usepackage{verbatim}
\usepackage{latexsym}

\begin{document}
\title{$V_{us}$ from $\pi$ and $K$ decay constants in full lattice QCD with physical $u$, $d$, $s$ and $c$ quarks}

\author{R.~J.~Dowdall}
\affiliation{DAMTP, University of Cambridge, Wilberforce Road, Cambridge CB3 0WA, UK}
\affiliation{SUPA, School of Physics and Astronomy, University of Glasgow, Glasgow, G12 8QQ, UK}
\author{C.~T.~H.~Davies}
\email[]{c.davies@physics.gla.ac.uk}
\affiliation{SUPA, School of Physics and Astronomy, University of Glasgow, Glasgow, G12 8QQ, UK}
\author{G. P. Lepage}
\affiliation{Laboratory of Elementary-Particle Physics, Cornell University, Ithaca, New York 14853, USA}
\author{C. McNeile}
\affiliation{Bergische Universit\"{a}t Wuppertal, Gaussstr.\,20, D-42119 Wuppertal, Germany}

\collaboration{HPQCD collaboration}
\homepage{http://www.physics.gla.ac.uk/HPQCD}
\noaffiliation

\date{\today}

\begin{abstract}
We determine the decay constants of the $\pi$ and $K$ mesons on 
gluon field configurations from the MILC collaboration 
including $u$, $d$, $s$ and $c$ quarks. 
We use three values of the lattice spacing and 
$u/d$ quark masses going down to the physical value. 
We use the $w_0$ parameter to fix the relative lattice spacing 
and $f_{\pi}$ to fix the overall scale. This allows us 
to obtain a value for $f_{K^+}/f_{\pi^+} = 1.1916(21)$. 
Comparing to the ratio of experimental leptonic decay 
rates gives 
$|V_{us}| = 0.22564(28)_{\mathrm{Br}(K^+)}(20)_{\mathrm{EM}}(40)_{\mathrm{latt}}(5)_{V_{ud}}$
and the test of unitarity 
of the first row of the Cabibbo-Kobayashi-Maskawa matrix: 
$|V_{ud}|^2+|V_{us}|^2+|V_{ub}|^2 - 1 = 0.00009(51)$. 

\end{abstract}

%\pacs{}
%\keywords{}

\maketitle

%%%%%%%%%%%%%%%%%%%%%%%%%%%%%%%%%%%%%%%%%%%%%%%%%%%%%%%%%%%%%%%%%
%
\section{Introduction}
\label{sec:intro}
The annihilation of a charged $\pi$ or $K$ meson to leptons 
via a $W$ boson is 
a `gold-plated' process whose rate can be determined very 
accurately from experiment. The decay width for pseudoscalar 
$P$ made of valence quarks $a\overline{b}$ is given by: 
\begin{equation}
\Gamma(P \rightarrow l \nu) = \frac{G_F^2|V_{ab}|^2}{8\pi}f_P^2m_l^2M_P\left(1-\frac{m_l^2}{M_P^2}\right)^2 
\label{eq:gamma}
\end{equation}
up to known electromagnetic corrections. Here $V_{ab}$ is 
the appropriate Cabibbo-Kobayashi-Maskawa matrix element
for coupling to the $W$ and $f_P$ is the pseudoscalar 
decay constant which parameterises the amplitude for the 
annihilation. $f_P$ can only be determined accurately from 
lattice QCD calculations. 

The experimental determination of 
$\Gamma(K^+ \rightarrow l \nu)/\Gamma(\pi^+ \rightarrow l \nu)$ 
can be converted to a result for the ratio of 
CKM element $\times$ decay constant 
for $K$ and $\pi$. 
Using experimental averages~\cite{pdg} 
$\Gamma(\pi^+ \rightarrow l \nu) = 3.8408(7) \times 10^7 s^{-1}$ and 
$\Gamma(K^+ \rightarrow l \nu) = 5.133(13) \times 10^7 s^{-1}$
gives:
\begin{equation}
\frac{|V_{us}|f_{K^+}}{|V_{ud}|f_{\pi^+}} = 0.27598(35)_{\mathrm{Br}(K^+)}(25)_{\mathrm{EM}}.
\label{eq:rat}
\end{equation} 
Here we have allowed for an electromagnetic correction to the ratio of 
widths given by $(1+\delta_{\mathrm{EM}})$ with 
$\delta_{\mathrm{EM}}=-0.0070(18)$\cite{flag, Cirigliano:2011tm}. 
The error in Eq.~(\ref{eq:rat}) from this correction is sizeable
but not as large as that from the $K^+$ branching fraction to $\mu \nu$ 
which dominates. The total error in determining the ratio of 
Eq.~(\ref{eq:rat}) from experiment is then 0.16\%.  

The electromagnetic correction means that $f_K$ and $f_{\pi}$
are defined as quantities in pure QCD without 
electromagnetic interactions. 
An accurate theoretical result from lattice QCD for 
$f_{K^+}/f_{\pi^+}$ then 
yields $|V_{us}|/|V_{ud}|$~\cite{Marciano:2004uf}. Since $V_{ud}$ is known accurately 
from nuclear $\beta$ decay, this gives $V_{us}$. 
The higher the accuracy on $V_{us}$ the more stringent the 
test of CKM first row unitarity we can do, since $V_{ub}$ 
is too small to contribute (at present) to this. 
Any deviations are indications of new physics and the 
more stringent the test, the higher the scale to which 
the new physics is pushed. 
 
State-of-the-art lattice QCD calculations have achieved 
errors below 1\% in $f_K/f_{\pi}$~\cite{Follana:2007uv, Durr:2010hr, Bazavov:2010hj, Aoki:2010dy, Laiho:2011np}, typically dominated by 
the systematic errors from extrapolation of the lattice 
results to the 
real-world continuum and chiral limits where the lattice 
spacing is zero and the $u/d$ quark masses take their 
physical values (equivalent to the $\pi$ meson mass taking 
its physical value). This means that significant improvements 
can be expected if we reduce discretisation errors, to 
make the continuum extrapolation more benign, and 
if we work with physical $u/d$ quark masses that obviate 
the need for a chiral extrapolation. It also means that 
comparison of different methods for arriving at 
a physical answer from lattice QCD are important in 
testing systematic error estimates. 

The MILC collaboration recently gave an analysis of 
$f_{K^+}/f_{\pi^+}$~\cite{Bazavov:2013cp} from lattice QCD on their `second-generation' 
gluon field configurations that include $u$, $d$, $s$ and $c$ 
quarks in the sea using the Highly Improved Staggered 
Quark (HISQ) formalism~\cite{Follana:2006rc} 
and a fully $\mathcal{O}(\alpha_sa^2)$ improved gluon action~\cite{Hart:2008sq}. 
They have ensembles with the average of the $u$ and $d$ quark 
masses down to the physical value. Their final error was 0.4\% 
on the decay constant ratio, dominated by errors from the 
extrapolation to zero lattice spacing. Their 
analysis~\cite{Bazavov:2013cp} concentrated 
on the ensembles with physical $u/d$ quark mass and the aim was 
to perform a single self-contained analysis that did not use additional 
information from, for example, chiral perturbation theory or determination 
of the lattice spacing using other quantities.  

In this paper we provide a new analysis with the most accurate 
result to date. To do this we use the same MILC ensembles with a 
completely independent analysis of 
meson correlation functions with high statistics. 
We include ensembles with heavier-than-physical 
$u/d$ quark masses and use chiral perturbation theory to pin down 
the point corresponding to physical light and strange quark masses.     
We also include very accurate information on the relative lattice 
spacings of the ensembles using the Wilson flow parameter, $w_0$~\cite{Borsanyi:2012zs}. 
This enables us to reduce the error on the decay constant ratio 
to below 0.2\% which is close to the error coming from experiment in Eq.~(\ref{eq:rat}). 

Section~\ref{sec:latt} describes the lattice calculation. The results and analysis, including 
a table of all our raw lattice values for meson masses and decay constants, 
are given in Section~\ref{sec:res}. This is followed by a discussion and 
conclusions in Sections~\ref{sec:discuss} and~\ref{sec:conclude} respectively. 

\section{Lattice Calculation}
\label{sec:latt}
\begin{table}
\caption{
Details of the MILC gluon field ensembles used in this paper~\cite{Bazavov:2010ru, Bazavov:2012uw}. 
$\beta=10/g^2$ is the $SU(3)$ gauge coupling and $L$ and $T$ 
give the length in the space and 
time directions for each lattice. 
$am_{\ell,sea},am_{s,sea}$ and $am_{c,sea}$ are the light (up and down taken to 
have the same mass), strange and charm sea quark masses in lattice units.
The ensembles 1, 2 and 3 will be referred to in the text as ``very coarse'', 
4, 5 and 6 as ``coarse'' and 7 and 8 as ``fine''. The number of 
configurations that we have used in each ensemble is given 
in the final column. We have used 16 time sources on every configuration.  
}
\label{tab:params}
\begin{ruledtabular}
\begin{tabular}{llllllll}
Set & $\beta$ &  $am_{\ell,sea}$ & $am_{s,sea}$ & $am_{c,sea}$ & $L/a\times T/a$ & $n_{\mathrm{cfg}}$ \\
\hline
1 & 5.80 &  0.013   & 0.065  & 0.838 & 16$\times$48 & 1020 \\
2 & 5.80 &  0.0064  & 0.064  & 0.828 & 24$\times$48 & 1000 \\
3 & 5.80 &  0.00235  & 0.0647  & 0.831 & 32$\times$48 & 1000 \\
\hline
4 & 6.00 &  0.0102  & 0.0509 & 0.635 & 24$\times$64 & 1052 \\
5 & 6.00 &  0.00507 & 0.0507 & 0.628 & 32$\times$64 & 1000 \\
6 & 6.00 &  0.00184 & 0.0507 & 0.628 & 48$\times$64 & 1000 \\
\hline
7 & 6.30 &  0.0074  & 0.0370  & 0.440 & 32$\times$96 & 1008 \\
8 & 6.30 &  0.0012 & 0.0363 & 0.432 & 64$\times$96 & 621 \\
\end{tabular}
\end{ruledtabular}
\end{table}

\begin{table}
\caption{
Values of the lattice spacing for the ensembles of 
Table~\ref{tab:params} in units of parameters 
$w_0$~\cite{Borsanyi:2012zs}, $t_0$~\cite{Luscher:2010iy} 
and $r_1$~\cite{Bazavov:2009bb}. The $r_1/a$ values 
were calculated by MILC and given in~\cite{Bazavov:2012uw}. 
}
\label{tab:latspace}
\begin{ruledtabular}
\begin{tabular}{llll}
Set &  $w_0/a$ & $\sqrt{t_0}/a$ & $r_1/a$ \\
\hline
1 & 1.1119(10) & 1.0249(5)  & 2.059(23)   \\
2 & 1.1272(7) & 1.0319(3)  & 2.073(13)   \\
3 & 1.1367(5) & 1.0357(2)  & 2.089(8)   \\
\hline
4 & 1.3826(11)  & 1.2389(5)  & 2.575(17)  \\
5 & 1.4029(9) & 1.2475(4) & 2.626(13)  \\
6 & 1.4149(6) & 1.2521(3) & 2.608(8)  \\
\hline
7 & 1.8869(39)  & 1.6515(16)  & 3.499(24)   \\
8 & 1.9525(20) & 1.6769(7) & 3.565(13)  \\
\end{tabular}
\end{ruledtabular}
\end{table}

\subsection{Meson Correlators}
Table~\ref{tab:params} gives the parameters for 
the MILC ensembles of gluon field configurations 
that we use here~\cite{Bazavov:2010ru, Bazavov:2012uw}. 
The table includes the values of 
sea quark masses in lattice units, where the $u$ 
and $d$ quarks are taken to have the same mass, 
$m_{\ell}=m_u=m_d$. 
The accurate determination of the lattice 
spacing will be discussed further below. Here we simply 
note that the `very coarse' lattices (sets 1, 2 and 3) have 
lattice spacing, $a\approx$ 0.15fm, the coarse lattices (sets 4, 5 
and 6) have $a\approx$ 0.12 fm and the fine lattices (sets 7 and 8) 
have $a\approx$0.09 fm. Thus the spatial volumes of the lattices 
are large: the sets with physical $m_l$ (sets 3, 6 and 8) 
are all larger than 4.8 fm on a side, with sets 6 and 8 
being larger than 5.5 fm on a side. 

On these ensembles we calculate light and $s$ quark propagators 
using the same HISQ action as used in the sea. The valence $\ell$ quarks 
are taken to have the same mass as those in the sea, the valence 
$s$ quarks are retuned slightly to correspond more closely 
to the physical value~\cite{Dowdall:2011wh}. The valence masses 
used are given in Table~\ref{tab:valence}. We use a unsmeared random wall 
source on each of 16 time sources per configuration for very high 
statistical accuracy~\cite{Dowdall:2011wh}. 

The propagators are combined to make meson correlation functions 
for $\pi$, $K$ and $\eta_s$ mesons. The $\eta_s$ is a fictitious 
$s\overline{s}$ meson that is not allowed to decay here because 
we do not include the disconnected pieces of the correlation 
function. Since it does not contain valence $u/d$ quarks it 
is a useful particle to study in lattice 
QCD~\cite{Davies:2009tsa, Dowdall:2011wh}. We include it here to 
provide more information to our fits about the meson mass dependence 
of the decay constants.  

\begin{figure}
\includegraphics[width=\hsize]{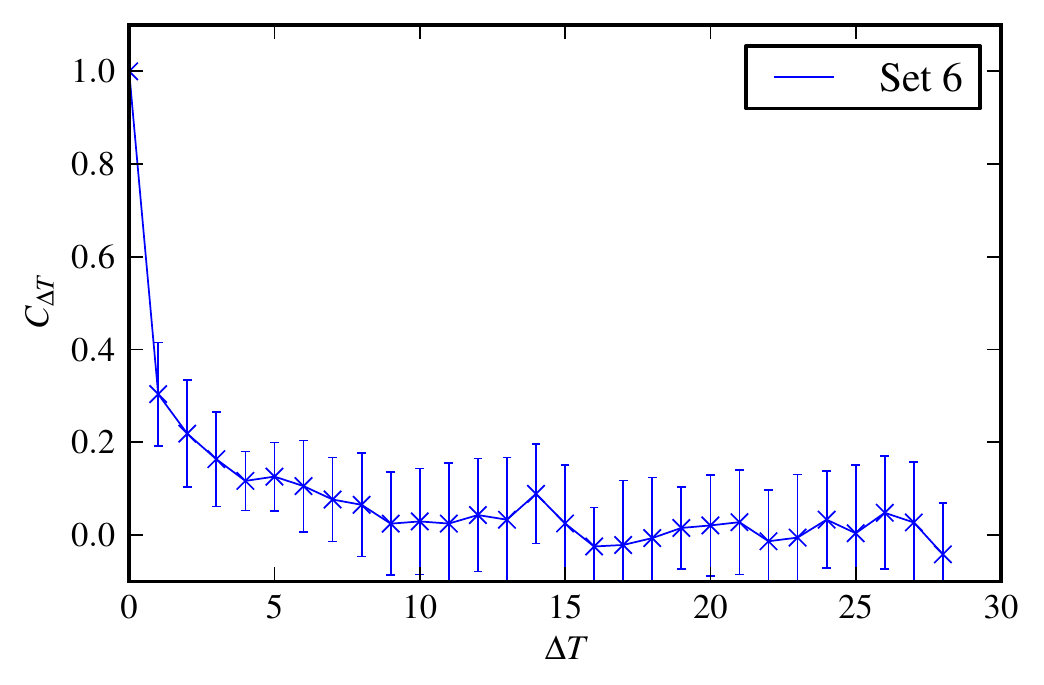}
\caption{\label{fig:autocorr} The autocorrelation function 
for the $\pi$ meson correlator at lattice time 20 on 
coarse set 6. $\Delta T = 1$ corresponds to adjacent configurations 
in the ordered list for the set which are 5 molecular dynamics time 
units apart. 
}
\end{figure}

We average over the 16 time sources to obtain a result for 
each configuration and then study the configuration-to-configuration 
correlations. 
An autocorrelation analysis was performed in~\cite{Dowdall:2011wh}
and plots of the autocorrelation function for $\pi$ and $\eta_s$ 
correlators given for the ensembles with $m_{\ell}/m_s = 0.1$ and 0.2. 
The autocorrelation function for the sets at physical $m_{\ell}/m_s$ 
show a decrease in correlation between 
configurations. We give an example in Fig.~\ref{fig:autocorr} 
for the $\pi$ meson correlator on 
coarse set 6. The other physical point ensembles show 
very similar behaviour. We bin over 2 adjacent configurations on 
sets 1, 2, 3 and 6 and bin over 4 adjacent configurations on all 
other sets. 

We then fit all three meson correlators
simultaneously as a 
function of time, $t$, between source and sink according to: 
\begin{eqnarray}
\label{eq:fitHISQ}
G_{\mathrm{meson}}(t)
 &=& \sum_{k=0}^{n_{\mathrm{exp}}} a_k( e^{-E_kt} + e^{-E_k(T-t)} )
\\
&-&
(-1)^{t/a}
\sum_{ko=0}^{n_{exp}} \tilde{a}_{ko} ( e^{-\tilde{E}_{ko}t} + e^{-\tilde{E}_{ko}(T-t)}  ). \nonumber
\end{eqnarray}
The oscillating piece is absent for $\pi$ and $\eta_s$ mesons because
the valence quark and antiquark have equal mass. 
We use Bayesian fitting methods~\cite{gplbayes, fitcode} so that the 
full effect of excitations in the spectrum can be included in the 
errors on the ground-state quantities that we are interested in, i.e. $a_0$ 
and $E_0$ for each meson. 
The simultaneous fit to all three mesons 
allows us to take into account the correlations between
the fit results for each meson in our subsequent analysis. 
Results are taken from 6 exponential fits (with an additional 
6 oscillating exponentials for the $K$). 

The $\pi$, $K$ and $\eta_s$ meson masses are given by the appropriate 
$E_0$ values from 
the fit above. The decay constant is determined 
from the corresponding amplitude, $a_0$, by
\begin{equation}
f_{ab} = (m_a + m_b) \sqrt{ \frac{2a_0}{E_0^3}   }
\label{eq:decayconstant}
\end{equation}
for meson with quark content $a\overline{b}$~\cite{Follana:2007uv}. 
This formula holds for Goldstone pseudoscalar mesons 
made of staggered quarks and follows from the existence of 
a partially conserved axial current relation in these formalisms. 
The decay constant is then absolutely normalised in
lattice QCD. 

Table~\ref{tab:valence} gives the results of our correlator fits 
for the decay constant and meson masses in lattice units on each 
ensemble. The errors are below 0.1\% in almost all cases. 

\begin{figure}
\includegraphics[width=\hsize]{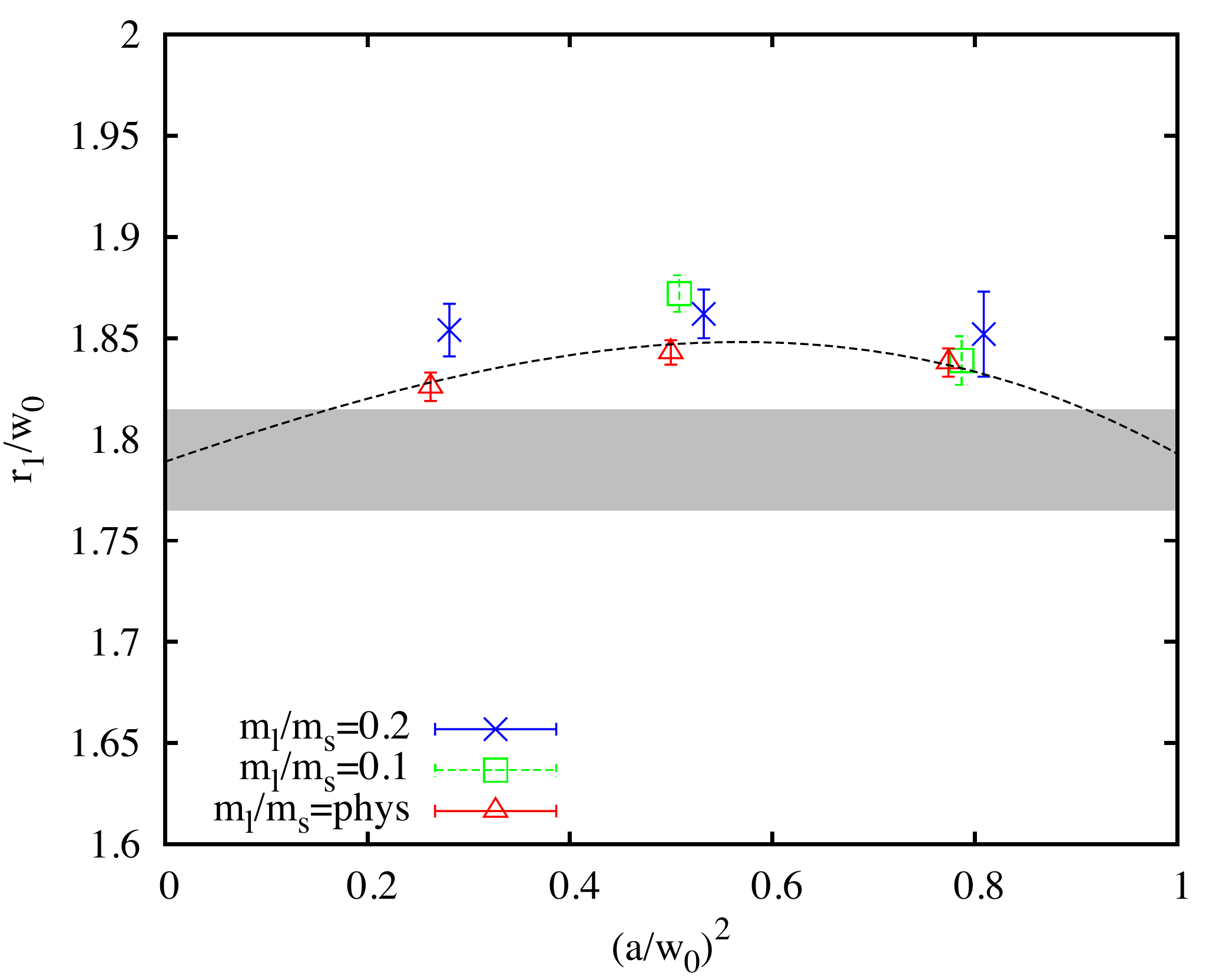}
\includegraphics[width=\hsize]{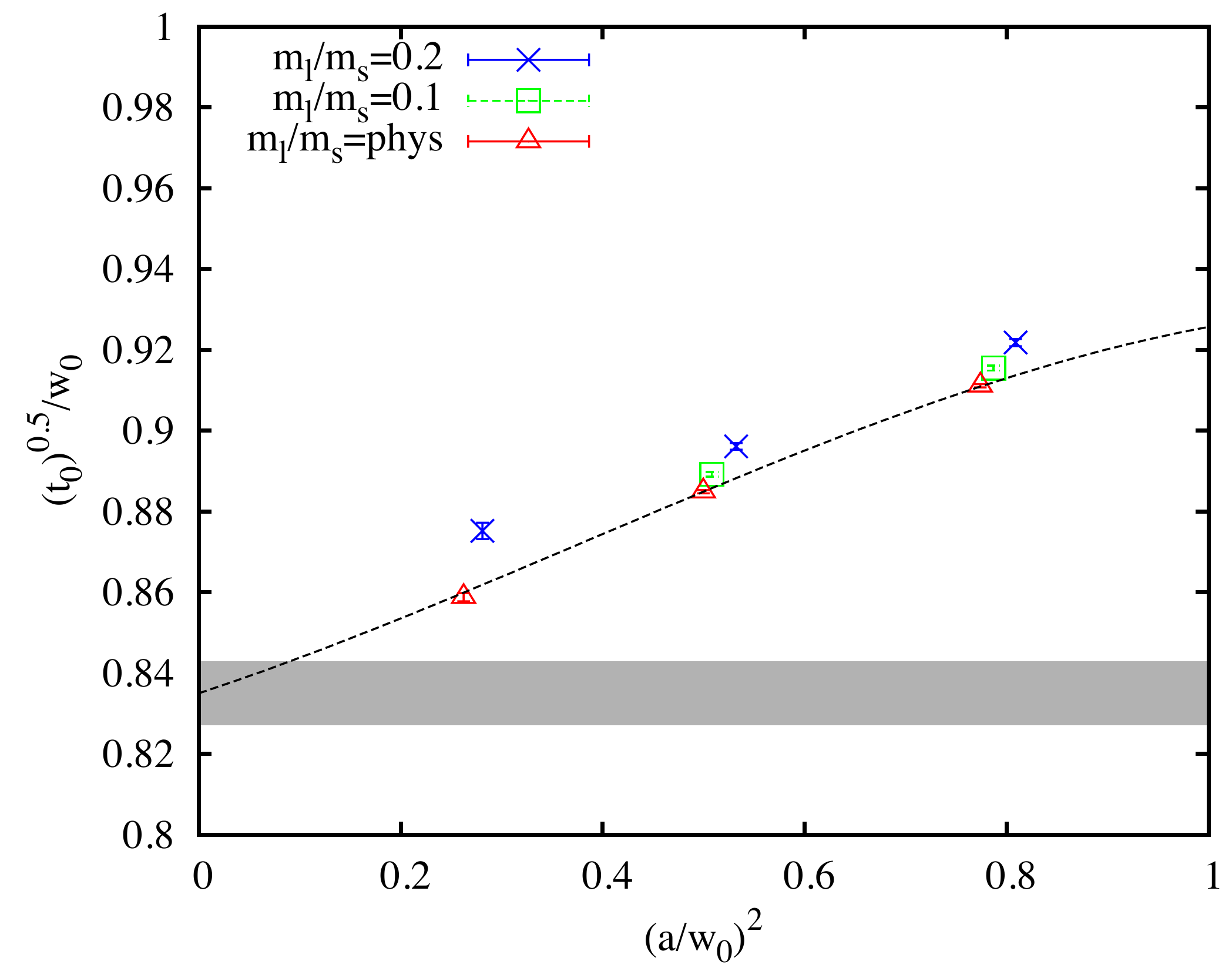}
\caption{\label{fig:r1w0} At the top the ratio $r_1/w_0$ plotted as 
a function of $a^2$ and below the ratio $\sqrt{t_0}/w_0$. 
The grey band gives the results of a 
simple polynomial fit to the $a^2$ and $m_{\ell}$ dependence 
as described in the text. The dashed line shows the 
fit evaluated at the physical value for $m_l/m_s$ of 0.036~\cite{pdg}.  
}
\end{figure}

\subsection{Lattice Spacing Determination}
It has recently been proposed that lattice spacings
be measured by smoothing the gluon field using a series 
of infinitesmal 
`smearing' steps~\cite{Luscher:2010iy, Borsanyi:2012zs}. 
This drives the gluon field towards a smooth 
renormalised field and gauge-invariant functions of this 
field, such as the pure gluon action density, become physical quantities.   
Parameters used to determine the lattice spacing can 
then be defined from the flow-time, $t$, dependence 
of such quantities. Ref.~\cite{Luscher:2010iy} defines $t_0$ 
from:
\begin{equation}
t^2\langle E \rangle |_{t=t_0} = 0.3
\end{equation}
where $\langle E \rangle$ is the expectation value of the 
gluon action density. 
The parameter $w_0$ is preferred in~\cite{Borsanyi:2012zs}, 
where $w_0$ is defined by: 
\begin{equation}
t \frac{d}{dt} t^2\langle E \rangle |_{t=w_0^2} = 0.3.
\end{equation}
$w_0$ should be less sensitive to small flow times where discretisation 
effects may be important. Both $w_0$ and $t_0$ can be determined by 
direct measurement on the gluon field and this makes them simpler to 
evaluate as well as typically more precise than parameters based 
on the heavy quark potential~\cite{Sommer:1993ce}. 
The heavy quark potential must be 
determined by fitting large Wilson loops as a function of (lattice) 
time and then, to extract parameters such as $r_1$~\cite{Bazavov:2010ru}, 
a further fit as a function of $r$ must be done to the potential. 

None of $w_0$, $t_0$ and $r_1$ can be simply related to any directly 
measurable experimental quantity and their physical value must be 
determined by a lattice QCD calculation of such a quantity. 
For example, $w_0$ and 
$t_0$ are determined from the mass of the $\Omega$ baryon 
in~\cite{Borsanyi:2012zs} and $r_1$ is determined from a basket of quantities including the 
$\Upsilon$ excitation energy and the decay constant of the $\eta_s$ 
meson in~\cite{Davies:2009tsa, Dowdall:2011wh}.  

Here we will use $w_0/a$ to determine the relative lattice spacing
between the ensembles and finally fix its value from $f_{\pi}$ 
in our analysis in Section~\ref{sec:res}. We give values for 
$w_0/a$ on each ensemble in Table~\ref{tab:latspace} 
and also, for comparison, of 
$\sqrt{t_0}/a$. These were obtained using the methods explained 
in~\cite{Borsanyi:2012zs}. We bin over 12 adjacent configurations 
(60 molecular dynamics time units on all sets except for set 8 
where it is 72 time units)
to remove the effects of autocorrelations in the results as 
assessed by a binning analysis.    

Fig.~\ref{fig:r1w0} compares the scales $w_0$, $t_0$ and $r_1$ by 
plotting $r_1/w_0$ and $\sqrt{t_0}/w_0$ as a function of 
$(a/w_0)^2$. We see that discretisation errors 
largely cancel between $r_1$ and $w_0$ since their ratio is very 
flat in $a^2$. There is a small variation with $m_{\ell}$. The errors 
here are dominated by statistical/fitting errors in $r_1$. 
In contrast $\sqrt{t_0}/w_0$ is much more precise but 
has relatively strong lattice spacing dependence, presumably 
from $\sqrt{t_0}$~\cite{Borsanyi:2012zs}. We carried 
out a simple
polynomial fit in $m_{\ell}/(10m_s)$ and 
$\alpha_s(\Lambda a)^2$, $(\Lambda a)^4$, $(\Lambda a)^6$ 
with $\Lambda$=0.6 GeV
for both ratios,
taking priors on the coefficients of 0.0(1.0).  
This gives the result $r_1/w_0 = 1.789(26)$ and 
$\sqrt{t_0}/w_0 = 0.835(8)$ in the continuum and physical light 
quark mass
limits, represented by the grey shaded bands in Fig.~\ref{fig:r1w0}. 
The result for $\sqrt{t_0}/w_0$ agrees well with the 
BMW-c result of 0.835(15)(7) using the Wilson clover 2-HEX action
in~\cite{Borsanyi:2012zs}. 

\section{Analysis and Results} 
\label{sec:res}

\begin{table*}
\caption{
Values for $\pi$, $K$ and $\eta_s$ masses and decay constants in lattice
units calculated for valence masses given in columns 2 and 3. Some of the results were previously 
given in~\cite{Dowdall:2011wh, McNeile:2012xh}. There are slight differences in 
some values with earlier results because different fit results were used. 
}
\label{tab:valence}
\begin{ruledtabular}
\begin{tabular}{lllllllll}
Set & $am_{{\ell},val}$ & $am_{s,val}$ & $aM_{\pi}$ & $af_{\pi}$ & $aM_{K}$ & $af_{K}$ & $aM_{\eta_s}$ & $af_{\eta_s}$ \\
\hline
1 & 0.013   & 0.0688 & 0.23644(15)& 0.11184(10) & 0.41214(24) & 0.12695(15) & 0.53350(17) & 0.14185(9) \\
  &         & 0.0641 &            &             & 0.40006(19) & 0.12585(10) & 0.51511(16) & 0.14009(7) \\
2 & 0.0064  & 0.0679 & 0.16614(7) & 0.10508(6)  & 0.39077(10) & 0.12265(4)  & 0.52798(9)  & 0.14027(4) \\
  &         & 0.0636 &            &             & 0.37948(10) & 0.12177(4)  & 0.51080(9)  & 0.13840(4) \\
3 & 0.00235 & 0.0628 & 0.10172(4) & 0.09938(6)  & 0.36557(8)  & 0.11837(4)  & 0.50656(6)  & 0.13720(2) \\
\hline
4 & 0.01044 & 0.0522 & 0.19158(9) & 0.09077(6) & 0.32789(11) & 0.10189(5) & 0.42358(11) & 0.11318(4) \\
5 & 0.00507 & 0.0505 & 0.13414(6) & 0.08452(5) & 0.30756(10) & 0.09788(4) & 0.41474(8)  & 0.11119(3) \\
6 & 0.00184 & 0.0507 & 0.08154(2) & 0.07990(3) & 0.29843(5)  & 0.09532(2) & 0.41478(4)  & 0.11065(2) \\
\hline
7 & 0.0074  & 0.0364 & 0.14062(10) & 0.06618(5) & 0.23919(11)& 0.07424(4) & 0.30871(10) & 0.08236(3) \\
8 & 0.0012  & 0.0360 & 0.05716(2)  & 0.05784(3) & 0.21855(5) & 0.06921(2) & 0.30480(4)  & 0.08053(2) \\
\end{tabular}
\end{ruledtabular}
\end{table*}

Table~\ref{tab:valence} gives our raw lattice results for the pseudoscalar
meson masses and decay constants in lattice units. In this section we use
these data to compute the dependence of $f_\pi$, $f_K$ and $f_{\eta_s}$ on the
quark masses and on the lattice spacing. This allows us to interpolate to the
physical values of the strange-quark and light-quark masses, and to
extrapolate to
zero lattice spacing, obtaining new predictions for the decay constants (as
well as the $\eta_s$~mass).  The differences between our most chiral
simulation data and our final  results are small since our simulation is very
close to physical. It is nevertheless important to model these corrections
accurately to optimise the precision of our final results.

Our analysis involves the following steps:
\begin{enumerate}
	\item Remove the lattice spacing by multiplying the masses and decay
	constants for the $\pi$, $K$, and $\eta_s$ by the values (with errors) 
	of $w_0/a$ from Table~\ref{tab:latspace}. We also apply an \emph{svd} cut
	to the data to guarantee that roundoff errors are not an issue when 
	inverting the covariance matrix for the $\chi^2$~function. This in effect
	triples the statistical errors.

    \item Fit the simulation results for $w_0 f_\pi$, $w_0 f_K$, $w_0
	f_{\eta_s}$, and $w_0^2 M_{\eta_s}^2$, together with the experimental
	result for $f_{\pi^+}$, as functions of the corresponding pion and kaon
	masses, and $w_0$. We take the functional dependence from one-loop
	partially-quenched chiral perturbation theory plus terms polynomial in
	$M_\pi^2$, $M_K^2$, and $a^2$. The fit gives a new value for $w_0$, which
	is largely determined by the experimental value for $f_\pi$ used in the fit.
	It also gives the functional dependence of the decay constants and the $\eta_s$
	mass on the quark masses, as specified by $M_\pi$ and $M_K$. Sea and 
	valence quark masses are specified separately. The same chiral formulas, 
	with the same couplings, are used for pions, kaons and $\eta_s$s; only
	the valence quark masses differ.

	\item Evaluate the best-fit functions for $f_K$ and $f_\pi$ at values
	of the pion and kaon masses appropriate for $f_{\pi^+}$ and 
	$f_{K^+}$. We set the $u$ and $d$ quark masses equal in our simulations. We 
	correct for this approximation through appropriate choices for the values of
	$M_\pi$ and $M_K$ used in our fit formulas to obtain our final results. At 
	the same time we correct for (small) electromagnetic corrections to the meson
	masses. (The decay constants, by definition, do not need electromagnetic
	corrections; these are included explicitly in Eq.~(\ref{eq:rat}).)
\end{enumerate}

In the rest of this section we elaborate on these steps, and survey
our results.

\subsection{Chiral Fit}
We fit our lattice results for the decay constants using a
formula drawn from partially-quenched chiral perturbation 
theory~\cite{Sharpe:2000bc} that has the following form:
\begin{equation}
	f_\mathrm{NLO}(x_a, x_b, x_\ell^\mathrm{sea}, x_s^\mathrm{sea}, L)
	+ \delta f_\chi + \delta f_\mathrm{lat}.
\label{eq:chiral}
\end{equation}
Here $f_\mathrm{NLO}$ is the result from chiral perturbation theory through
1-loop order, in a finite volume of size $L$ on a side; 
and $\delta f_\chi$ and $\delta f_\mathrm{lat}$ are corrections for 
higher-order chiral contributions and nonzero lattice-spacing errors, respectively.
We specify valence and sea quark masses through the dimensionless parameters
$x_a$, $x_b$, \emph{etc.} where, for example, 
a light-quark with mass $m_\ell = (m_u+m_d)/2$ would correspond to
\begin{equation}
	x_\ell = \frac{{M}_{0,\pi}^2}{16\pi^2 f_0^2}.
	\label{eq:xl}
\end{equation}
Here $f_0$ is fixed to the standard value 131.5 MeV, 
close to the experimental result for $f_\pi$. 
$M_{0,\pi}$ is the bare pion mass, obtained by subtracting the 1-loop chiral
correction from masses measured in the simulation (Table~\ref{tab:valence}). 
Using bare meson  masses corrects for (negligible) finite-volume errors in the
masses. The  $s$-quark parameter is given by
\begin{equation}
	x_s = \frac{2 M_{0,K}^2 - M_{0,\pi}^2}{16\pi^2 f_0^2}
	\label{eq:xs}
\end{equation}
where $M_{0,K}$ is the bare mass coming from the kaon masses measured in the
simulation. The same formula is used for each of the three mesons we study,
changing only the valence masses:
\begin{align}
	f_\pi &\longleftrightarrow f_\mathrm{NLO}(x_\ell, x_\ell, x_\ell^\mathrm{sea}, x_s^\mathrm{sea}, L)
	+ \delta f_\chi + \delta f_\mathrm{lat} \nonumber \\
	f_K &\longleftrightarrow 
	f_\mathrm{NLO}(x_\ell, x_s, x_\ell^\mathrm{sea}, x_s^\mathrm{sea}, L)
	+ \delta f_\chi + \delta f_\mathrm{lat} \nonumber \\
	f_{\eta_s} &\longleftrightarrow 
	f_\mathrm{NLO}(x_s, x_s, x_\ell^\mathrm{sea}, x_s^\mathrm{sea}, L)
	+ \delta f_\chi + \delta f_\mathrm{lat}
\end{align}

\begin{table*}
\caption{Finite-volume corrections, $\Delta_\mathrm{vol} f$,
to simulation results for the meson decay constants. 
Errors on the finite-volume correction come from our fit and 
are correlated between ensembles and between $\pi$ and $K$. 
Also listed for each ensemble are the lattice 
spacing~$a$ (after determination of $w_0$ which gives the error shown, 
correlated between ensembles), the ratio of valence strange to light-quark mass $m_s/m_\ell$,
the spatial dimension of the lattice~$L$, and the pion and 
kaon masses (with their statistical errors from Table~\ref{tab:valence}).}
\label{tab:finite-volume}
\begin{ruledtabular}
\begin{tabular}{ccccccccc}
      $a$     &     $m_s/m_\ell$  &   $L$    &     $M_\pi L$    
      &       $M_\pi$      &      $M_K$  &  $\Delta_\mathrm{vol} f_\pi$  
      &    $\Delta_\mathrm{vol} f_K$  & $\Delta_\mathrm{vol} f_{\eta_s}$ \\ \hline
   0.1543(8)\,fm &    5.3 &   2.5\,fm &    3.8   &  302.4(2)\,MeV & 527.1(3)\,MeV &  1.24(23)\% & 0.50(9)\% & 0.10(0)\% \\
   0.1522(8)\,fm &   10.6 &   3.7\,fm &    4.0   &  215.5(1)\,MeV & 506.8(1)\,MeV &  0.38(7)\%  & 0.12(2)\% & 0.00(0)\% \\
   0.1509(8)\,fm &   26.7 &   4.8\,fm &    3.3   &  133.0(1)\,MeV & 477.9(1)\,MeV &  0.43(8)\%  & 0.13(2)\% & 0.00(0)\% \\
   \hline 
   0.1241(7)\,fm &    5.0 &   3.0\,fm &    4.6   &  304.5(1)\,MeV & 521.2(2)\,MeV &  0.37(7)\%  & 0.14(3)\% & 0.01(0)\% \\
   0.1223(6)\,fm &   10.0 &   3.9\,fm &    4.3   &  216.5(1)\,MeV & 496.4(2)\,MeV &  0.24(5)\%  & 0.08(1)\% & 0.00(0)\% \\
   0.1212(6)\,fm &   27.6 &   5.8\,fm &    3.9   &  132.7(0)\,MeV & 485.7(1)\,MeV &  0.15(3)\%  & 0.05(1)\% & 0.00(0)\% \\
   \hline
   0.0907(5)\,fm &    4.9 &   2.9\,fm &    4.5   &  306.1(2)\,MeV & 520.6(2)\,MeV &  0.41(8)\%  & 0.16(3)\% & 0.02(0)\% \\
   0.0879(4)\,fm &   30.0 &   5.6\,fm &    3.7   &  128.4(0)\,MeV & 490.8(1)\,MeV &  0.21(4)\%  & 0.07(1)\% & 0.00(0)\% \\
\end{tabular}
\end{ruledtabular}
\end{table*}
In our fits, we use very broad priors for the chiral parameters in
$f_\mathrm{NLO}$\,---\,10--100~times wider than the final errors\,---\,so
these have no impact on the fit. We also introduce a new parameter that
multiplies the finite-volume correction in $f_\mathrm{NLO}$. This allows
our fit to correct (crudely) for  finite-volume corrections from higher orders
in chiral perturbation theory. We set its prior to~$1\pm0.33$. Finite-volume
corrections are quite small on almost all of the ensembles, as is evident from 
Table~\ref{tab:finite-volume} which lists corrections for the decay constants.
(The $\eta_s$ mass has very small corrections, 
similar in magnitude to those for $f_{\eta_s}$). The finite-volume corrections 
agree at the level of the errors we have 
with the range of those calculated in~\cite{Colangelo:2005gd}, as well as with 
the finite-volume analysis in~\cite{Bazavov:2013cp}. 

The square of the $\eta_s$ mass is fit with an analogous formula of the form
\begin{equation}
	M_\mathrm{NLO}^2(x_a, x_b, x_\ell^\mathrm{sea}, x_s^\mathrm{sea}, L)
	+ \delta M^2_\chi + \delta M^2_\mathrm{lat},
\end{equation}
with $x_a=x_b=x_s$.

\subsection{Higher-Order Corrections}
We include terms beyond one-loop order in chiral perturbation theory by 
adding a correction of the form
\begin{align}
	\delta f_\chi &\equiv c_{2a} (x_a+x_b)^2 + c_{2b} (x_a-x_b)^2
	\nonumber \\
		&+ c_{2c} (x_a+x_b)(2x_\ell^\mathrm{sea} + x_s^\mathrm{sea})
		+ c_{2d} (2x_\ell^\mathrm{sea} + x_s^\mathrm{sea})^2
	\nonumber \\
		&+ c_{2e} (2x_\ell^\mathrm{sea\,2} + x_s^\mathrm{sea\,2}) 
	\nonumber \\
		&+ c_{3a} (x_a + x_b)^3 + c_{3b} (x_a+x_b)(x_a-x_b)^2 
	\nonumber \\
		&+ c_{3c}(x_a+x_b)^2 (2x_\ell^\mathrm{sea} + x_s^\mathrm{sea}) 
	\nonumber \\
		&+c_{4} (x_a+x_b)^4 + c_5 (x_a+x_b)^5 + c_6 (x_a+x_b)^6
	\nonumber \\
\label{eq:fchi}
\end{align}
where we take priors of~$0\pm 1$ for each parameter $c_j$. We only
keep higher-order terms that might be significant given the precision
of our  simulation data. In fact, we obtain
an excellent fit and almost identical results (to within a quarter of a 
standard deviation) when we keep only the quadratic terms. 
We include an analogous correction, $\delta M^2_\chi$,
for the square of the $\eta_s$~mass.

We also correct for the nonzero lattice-spacing using
\begin{equation}
	\delta f_\mathrm{lat} \equiv \sum_{n=1}^4 d_n \left(
	\frac{a\Lambda_\mathrm{QCD}}{\pi} \right)^{2n}
	\label{eq:dn}
\end{equation}
where $d_n$ is allowed to depend upon the quark masses,
\begin{align}
	d_n &= d_{n,0} + d_{n,1a} (x_a+x_b) + 
	d_{n,1b} (2x_\ell^\mathrm{sea} + x_s^\mathrm{sea}) 
	\nonumber \\
	 &+ d_{n,1c} (x_a+x_b)^2,
\end{align}
and again the coefficients have priors~$0\pm1$. We get excellent fits
and almost identical answers without allowing for mass dependence, but 
we are conservative and include this possibility, since it could 
arise from taste-changing effects~\cite{Aubin:2003uc, Davies:2009tsa}, thereby increasing
our final errors by about half a standard deviation. 

Eq.~(\ref{eq:dn}) is an expansion in the QCD scale $\Lambda_\mathrm{QCD}$
divided by the ultraviolet cutoff, $\pi/a$, for the lattice. The QCD scale is
of order~500\,MeV to~1\,GeV. This is confirmed by the Empirical Bayes criterion~\cite{gplbayes}
which shows that our data imply a scale of about~600\,MeV. In our analysis we
use a conservative value,~1.8\,GeV, to ensure that nonzero lattice-spacing
errors are not underestimated.

\subsection{Isospin Violation and Electromagnetism} 
We need to determine the $x_\ell$ and $x_s$ values corresponding
to physical pion and kaon masses if we are to use our formulas to extract
physical values for the decay constants and the $\eta_s$~mass.
The correct pion and kaon masses come from
experiment, but there are two complications that result from simplifications
in the  simulations. The first is that the simulation does not include
electromagnetism. The second is that $m_u = m_d$ in the
simulation, while in reality $m_u = 0.48(10)\,m_d$~\cite{pdg}.

The most appropriate pion mass for $f_{\pi^+}$ is the 
neutral-pion mass (134.9766(6) MeV~\cite{pdg}). 
All $\pi$~mesons would have this mass in a world without 
electromagnetism\,---\,our simulations, for example\,---\,up
to very small (quadratic) corrections from the $u\mathrm{-}d$~mass difference. 
These corrections are estimated at 0.32(20)\,MeV for $M_{\pi^+}$ 
in~\cite{Amoros:2001cp}. For our purposes, it is sufficient to take 0.32\,MeV as the 
uncertainty in the pion mass, and ignore the distinction between charged
and neutral pions:
\begin{equation}
\label{eq:mpi}
	M_{\pi}^\mathrm{phys} = 134.98(32)\,\mathrm{MeV}
\end{equation}

This pion mass corresponds in our simulation to a light-quark mass of $m_\ell
= (m_u + m_d)/2$. The corresponding kaon mass is one for an
$s\overline\ell$~meson. This is the root-mean-square average of the $K^+$ and
$K^0$ masses with additional small corrections for electromagnetism:
\begin{align}
(M_{K}^\mathrm{phys})^2 \equiv\,&\frac{1}{2}\left[(M_{K^+}^2 + M_{K^0}^2) \right.
\nonumber \\
&\left. - (1+\Delta_E)(M_{\pi^+}^2-M_{\pi^0}^2) \right].
\end{align}
$\Delta_E$ would be zero if electromagnetic effects in the $K$ system 
mirrored those of the $\pi$. In fact it is closer to 1. Recent 
lattice calculations~\cite{Basak:2012zx, Portelli:2012pn, Blum:2010ym} 
that include electromagnetic effects give 
values in the region 0.6-0.7. We take $\Delta_E=0.65(50)$ to 
conservatively encompass these results and this gives 
\begin{equation}
\label{eq:mk}
	M_{K}^\mathrm{phys}=494.6(3)\,\mathrm{MeV}.
\end{equation}

Tuning the pion mass to $M_\pi^\mathrm{phys}$ and the kaon mass to
$M_K^\mathrm{phys}$ in our fits sets the strange-quark mass to its physical
value, and the light-quark  mass to the average~$m_\ell$ of the  $u$~and
$d$~masses. This light-quark mass is correct, to within our errors, for the
valence quarks in the pion, and for sea quarks in all three mesons.

This tuning is not correct, however, for the $K^+$'s valence light-quark,
which is a $u$~quark, with mass~$0.65(9)\,m_\ell$. This difference
produces a small but significant downward shift in~$f_{K^+}$.  To compute the
corrected $K^+$~decay constant, we evaluate our fit 
formulas with a pion mass given by
$\sqrt{0.65(9)} M_\pi^\mathrm{phys}$,  while adjusting the kaon mass so
that $2M_K^2-m_\pi^2$ is unchanged (to leave the  $s$-quark mass unchanged).
These adjustments are made  only for the valence-quark masses in the $K^+$;
the  valence-quark masses in the pion and~$\eta_s$, as specified
by~$M_\pi^\mathrm{phys}$ and~$M_K^\mathrm{phys}$, are left unchanged, as are
the  sea-quark masses in each of the mesons.

\subsection{Fit Results}
\begin{figure}
\includegraphics[width=\hsize]{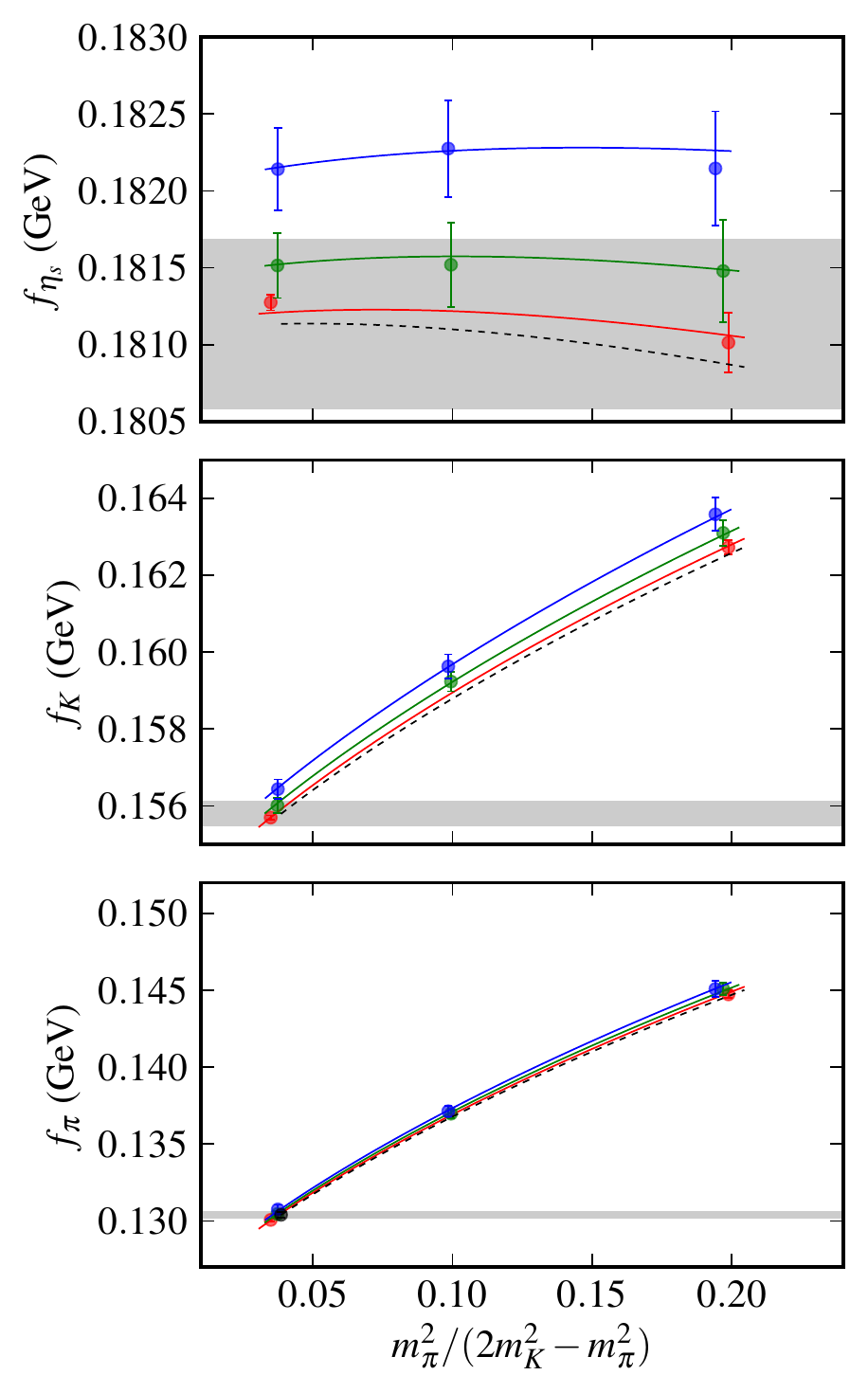}
\caption{\label{fig:fps} Fit results for the  $\pi$, $K$, and
$\eta_s$ decay constants as functions of the light-quark mass for three
different lattice spacings: 0.15\,fm (top/blue),  0.12\,fm (middle/green),
and 0.09\,fm (bottom/red). The data shown are  from Table~\ref{tab:valence},
with corrections for errors in the $s$~masses,
and for finite-volume errors. 
The lines show our fit  with the best-fit values of the fit parameters.
The dashed line is the $a=0$ extrapolation, and the gray band shows our continuum
results at the physical light quark mass point with $m_\ell = (m_u+m_d)/2$. 
The current experimental result for $f_{\pi^+}$ is also shown (black point).
Note that the three plots are against very different scales in the vertical
direction: the range covered in the $f_\pi$~plot is 10~times larger than 
that covered in the $f_{\eta_s}$~plot.}
\end{figure}

We fit $w_0$ times each of the decay constants and each $\eta_s$~mass 
in Table~\ref{tab:valence} to the 
formulas above, as functions of the pion and kaon masses and~$w_0$. We
also fit the experimental value for $f_{\pi^+} = 130.4(2)$\,MeV to our formula
evaluated at the physical pion and kaon masses, Eqs.~(\ref{eq:mpi}, \ref{eq:mk})). 
These fits are all done simultaneously using the same parameters for the fit
functions in each case, and including the correlations between 
$\pi$, $K$ and $\eta_s$ results discussed in Section~\ref{sec:latt}. 

The results for the decay constants, as a function of the light-quark mass,
are shown in  Figure~\ref{fig:fps}. For each
decay constant we show fit results and simulation  data for each
of our three lattice
spacings. The dashed line shows the continuum extrapolation, while
the gray band shows our final results extrapolated to zero lattice spacing
and the physical light quark mass limit (with the light-quark mass equal to the 
$u\mathrm{-}d$~average). The fit is excellent with a~$\chi^2$ per degree
of freedom of~0.42 (p-value 0.99), fitting 39~pieces of data. There are 
61 parameters, each with a Bayesian prior. The final results are:
\begin{align}
	f_\pi &= 130.39(20)\,\mathrm{MeV}  
		& f_{K^+}/f_{\pi^+} &= 1.1916(21) \nonumber \\
	f_{K^+} &= 155.37(34)\,\mathrm{MeV} 
		& f_{\eta_s}/M_{\eta_s} &= 0.2631(11) \nonumber \\
	f_{\eta_s} &= 181.14(55)\,\mathrm{MeV} 
		& M_{\eta_s}^2/(2M_K^2 - M_\pi^2) &= 1.0063(64)\nonumber \\
	M_{\eta_s} &= 688.5(2.2)\,\mathrm{MeV} 
		& f_{\eta_s}/(2f_K-f_\pi) &= 0.9997(17) \nonumber \\
	w_0 &= 0.1715(9)\,\mathrm{fm}
	\label{eq:results}
\end{align}
Clearly the result for $f_{\pi}$ contains no new 
information beyond the input value from experiment that 
was included as a fit parameter. 
The $K^+$~results here are adjusted to correct the valence light-quark mass,
as discussed above. We find that the $K^+$~decay constant is 0.27(7)\%~lower
than the decay constant for a kaon whose valence light-quark's mass equals the
$u\mathrm{-}d$~average mass.

\begin{figure}
\includegraphics[width=\hsize]{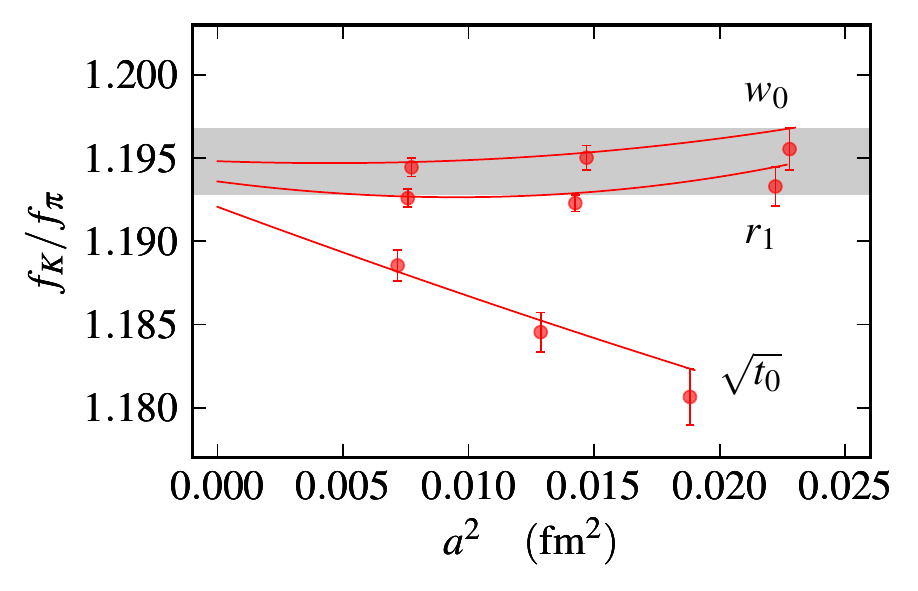}
\caption{\label{fig:a2dep} Fit results for $f_K/f_\pi$ evaluated
at the physical light quark mass limit, with $m_\ell = (m_u+m_d)/2$,  for different lattice spacings. 
The data shown are  from Table~\ref{tab:valence},
with corrections for errors in the quark masses,
and for finite-volume errors. The top curve
and data are from our analysis using $w_0$~to set the lattice spacing;
the middle results are from our analysis using $r_1$ instead of $w_0$;
and the bottom results are from our analysis using~$\sqrt{t_0}$. 
The gray band shows the final result from the $w_0$~analysis.
}
\end{figure}

Error budgets for several of our results are presented in
Table~\ref{tab:ebudget}. Our fits are unchanged if we include additional
higher-order chiral or $a^2$~corrections, beyond what is discussed above. 
Omitting results from any one of our
configuration sets shifts the mean values by no more than one standard
deviation and usually much less. Omitting results  from the smallest lattice
spacing (0.09\,fm) gives the same mean values but with standard deviations
that are 2.5~times larger. Omitting the most chiral results ($m_s/m_l > 25$)
shifts the means by about $1/3$~of a standard deviation and increases the
standard deviation by~50\%. These last two tests are evidence 
that our $a^2$~and chiral extrapolations are stable and robust.

As a check of the `statistical+svdcut' elements of the error 
budget we repeated the analysis using correlator 
results binned over many more adjacent configurations. We used 
a bin size of 16 corresponding to 80 molecular dynamics time 
units (64 or 96 on set 8 depending on stream). 
This gives the same final result within 
$0.5\sigma$ and the same errors.     

\begin{table}
\caption{
Sources of uncertainty in the final results (Eq.~(\ref{eq:results})) 
for the $K^+$~decay constant,
the ratio of~$K^+$ to~$\pi^+$ decay constants, the $\eta_s$ mass, and
the Wilson flow parameter~$w_0$.
}
\label{tab:ebudget}
\begin{ruledtabular}
\begin{tabular}{rlllll}
& $f_{K^+}$ & $f_{K^+}/f_{\pi^+}$ & $m_{\eta_s}$ & $w_0$ \\ \hline
statistics $+$ \emph{svd} cut & 0.13\% & 0.13\% & 0.28\% & 0.26\% \\
chiral extrapolation &  0.03 & 0.03 & 0.04 & 0.15 \\
$a^2\to0$ extrapolation & 0.10 & 0.10 & 0.15 & 0.27  \\ 
finite volume correction & 0.01 & 0.01 & 0.01 & 0.02 \\
$w_0/a$ uncertainty & 0.02 & 0.02 & 0.02 & 0.28 \\
$f_{\pi^+}$ experiment & 0.13 & 0.03 & 0.07 & 0.19 \\
$m_u/m_d$ uncertainty & 0.07 & 0.07 & 0.00 & 0.00 \\ \hline
Total & 0.22\% & 0.18\% &  0.33\%  & 0.54\% 
\end{tabular}
\end{ruledtabular}
\end{table}

The $a^2$~variation of our simulation results is quite small (1--2~standard
deviations) across our  entire range of lattice spacings. This is illustrated
in Fig.~\ref{fig:a2dep} where we show simulation results for $f_K/f_\pi$ in
the physical light quark mass limit for our three lattice spacings (top curve); the gray band
is the $a=0$~result. This behavior is in
marked contrast with what we obtain if we set the lattice spacing
using~$\sqrt{t_0}$ (bottom curve). The two methods agree to within
1.3~standard  deviations when extrapolated to~$a=0$, but the variation
with~$a^2$ in
the $\sqrt{t_0}$~analysis is much larger. This agrees with the findings 
of~\cite{Borsanyi:2012zs} that $\sqrt{t_0}$ has larger discretisation 
errors than $w_0$ when compared to hadronic quantities. 
We have also redone our analysis
using $r_1$ (middle curve). These results are similar to those from 
the $w_0$ analysis, and give an
extrapolated value that agrees with that analysis to within half a standard
deviation. These two analyses also give:
\begin{align}
	\sqrt{t_0} &=  0.1420(8)\,\mathrm{fm}    \nonumber \\
	r_1 &=    0.3112(30)\,\mathrm{fm}  
\label{eq:results2}
\end{align}
We use quite broad priors for $w_0$, $\sqrt{t_0}$, and $r_1$ in our fits: 
0.1755(175), 0.1400(140), and 0.3150(320), respectively. They have little 
effect on the fit results.

Finally, we give the Gasser-Leutwyler low-energy constants from the 
NLO term in our chiral fit. These are evaluated at scale $M_{\eta}$ 
and given in units of $10^{-3}$.
\begin{align}
	L_4 &= 0.36(34)  
		& L_6 &= 0.32(20) \nonumber \\
	L_5 &= 2.00(25)
		& L_8 &= 0.77(15) \nonumber \\
	2L_6-L_4 &= 0.28(17)
		& 2L_8-L_5 &= -0.46(20) \nonumber \\
	\label{eq:resLi}
\end{align}
Values agree well with other chiral analyses, for example 
the MILC analysis on configurations including 
$u$, $d$ and $s$ asqtad sea quarks~\cite{Bazavov:2009tw}. 
The errors on the low-energy constants 
reflect the fact that we allow for higher 
order terms beyond NLO chiral perturbation theory in our fits.

Chiral extrapolation contributes much less
to our error budget here than in our previous analyses. This is expected because
we have lattice results for $m_\ell$ very close to the physical mass; indeed,
$m_\ell$ is actually slightly below the physical mass, so we are interpolating.
We checked our chiral extrapolation in several ways:
\begin{itemize}
	\item We replaced the fixed value of $f_0$ that sets the chiral scale in Eqs.~(\ref{eq:xl}) 
	and (\ref{eq:xs}) with the floating parameter that corresponds to 
	$f_{\pi}$ in the chiral limit. Any changes should be absorbed by 
	the higher-order mass-dependent terms in the chiral fit, so this
	tests whether we have included enough higher order terms. 
	Changing $f_0$ in this way had negligible effect on our final 
	answers (around $\sigma/20$).  

	\item We replaced SU(3) chiral perturbation theory with SU(2) chiral 
	perturbation theory, where the chiral parameters for pions, kaons
	and the $\eta_s$ are allowed to differ. Unlike in the SU(3) case, it 
	is possible to fit our data with the SU(2) theory expanded only through
	next-to-leading order (NLO); results are the same as above
	to within~$1\sigma$, with slightly smaller errors. We prefer to include 
	analytic terms from next-to-next-to-leading order (NNLO) and above to
	ensure that we have not underestimated our errors. This again gives 
	results that agree with those in Eq.~(\ref{eq:results}) (to better 
	than $0.5\sigma$) but now with slightly larger errors~(by $0.1\sigma$).
        See~\cite{Bazavov:2009ir} and~\cite{Bazavov:2009tw} for a comparison of SU(2) and SU(3) chiral fits with similar conclusions for results with asqtad $u$, $d$ and $s$ sea quarks.  

	\item We repeated our analysis using 
	staggered chiral perturbation theory through one-loop~\cite{Aubin:2003uc} 
	supplemented by higher order terms in meson masses 
	and discretisation effects, as given earlier in Eqs.~(\ref{eq:chiral}),
	(\ref{eq:fchi}) and (\ref{eq:dn}).
	Staggered chiral perturbation theory explicitly incorporates 
	discretisation effects that arise when using staggered 
	quarks because of the multiple tastes of mesons that 
	can appear in loop terms in the chiral expansion. 
	The standard chiral logarithm terms are modified to include 
	multiple tastes and 
	in addition there are `hairpin' correction terms that 
	also depend on taste-splittings. The final effect 
	from these two terms is very benign, even at physical 
	quark masses. 
	We find that final results from our fits differ by less than 
	0.5$\sigma$ from the results given above in Eq.~(\ref{eq:results}).  
\end{itemize}
These tests give us confidence that our estimates of the errors 
due to our chiral fits are reliable.

\section{Discussion}
\label{sec:discuss}

\begin{figure}
\includegraphics[width=\hsize]{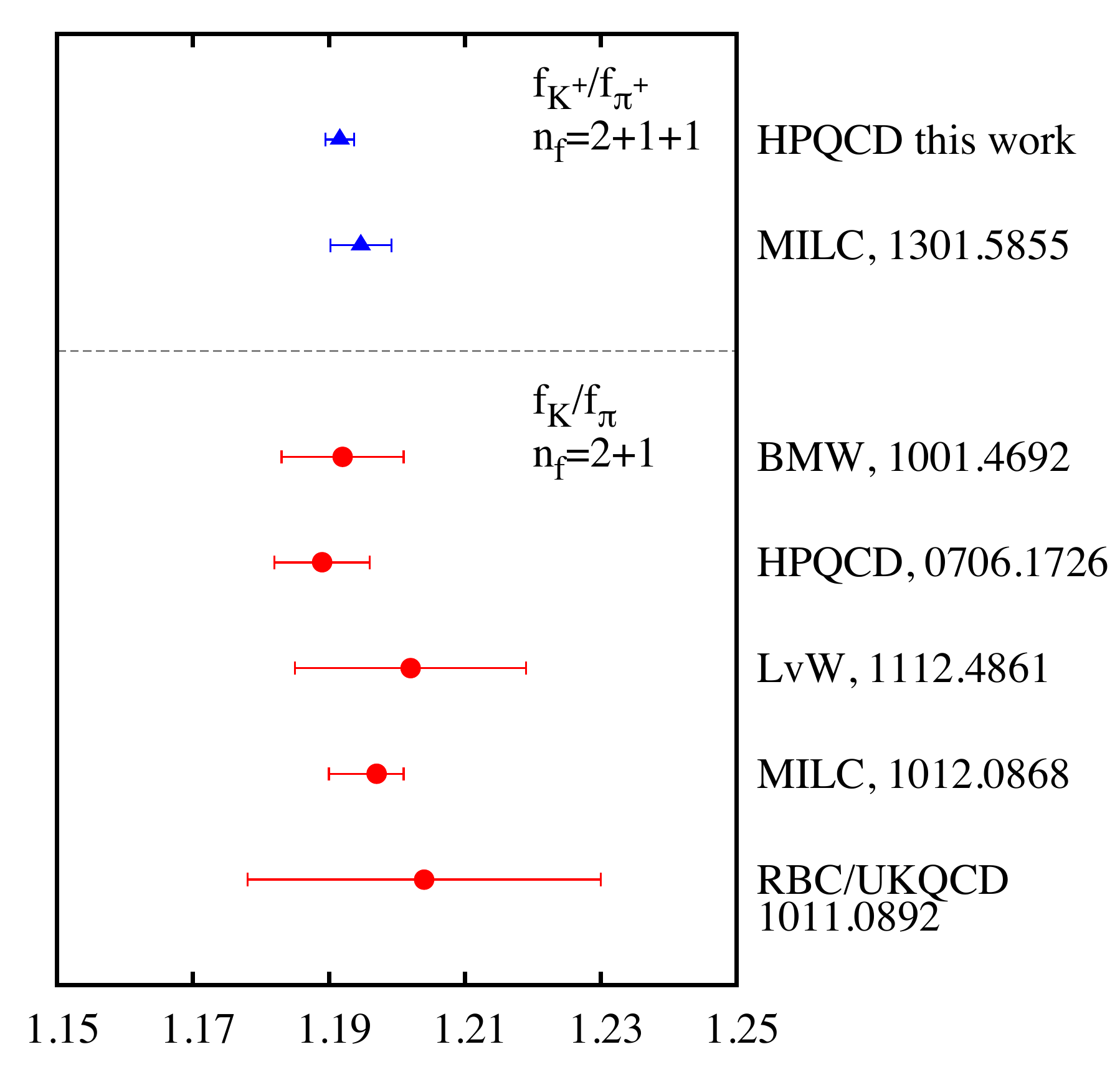}
\caption{\label{fig:fkpicomp} A comparison of lattice 
QCD results for the ratio of $K$ to $\pi$ decay constants. 
The top two values (filled blue triangles) are 
for $f_{K^+}/f_{\pi^+}$ including $u$, $d$, $s$ and 
$c$ sea quarks - results from this paper and from~\cite{Bazavov:2013cp}. 
The lower values (red squares) include $u$, $d$ 
and $s$ sea and typically do not distinguish $f_{K^+}$ 
from $f_K$~\cite{Durr:2010hr, Follana:2007uv, Laiho:2011np, Bazavov:2010hj, Aoki:2010dy}. 
}
\end{figure}

Eq.~(\ref{eq:results}) lists a number of outputs from our 
analysis. 
The key result is 
that for $f_{K^+}$ and in particular the ratio 
$f_{K^+}/f_{\pi^+}$ needed to make use of Eq.~(\ref{eq:rat}).
This is obtained with an error of 0.18\%. 

Fig.~\ref{fig:fkpicomp} compares our new result 
for $f_{K^+}/f_{\pi^+}$ to earlier 
values on $n_f=2+1+1$ ($u$, $d$, $s$ and $c$ sea 
quarks)~\cite{Bazavov:2013cp} 
and $n_f=2+1$ ($u$, $d$, $s$ sea quarks) configurations 
~\cite{Follana:2007uv, Durr:2010hr, Bazavov:2010hj, Aoki:2010dy, Laiho:2011np}.
There is good agreement with earlier results within 
their larger error bands. Typically the results on 
$n_f=2+1$ did not distinguish between $f_{K^+}$ and $f_K$ 
because the errors were not small enough to see this 
difference. The difference that we see, 0.27(7)\%, is 
in agreement with that expected from chiral perturbation 
theory (0.21(6)\%~\cite{Cirigliano:2011tm}). It is also 
in agreement with a simple-minded argument 
assuming that the ratio of $f_{\eta_s}/f_K$ 
depends linearly on $M^2_{\pi}$ between 0 and $M_{\eta_s}^2$. 

Our result for $f_{K^+}/f_{\pi^+}$ is more accurate than MILC's recent 
analysis~\cite{Bazavov:2013cp} based on the physical 
point ensembles because we have 
included additional information: accurate 
relative lattice spacing values, fits to decay constants 
at heavier sea and valence quark masses, fits to $\eta_s$ 
masses and decay constants, and
chiral perturbation theory to relate all of these fits 
at all of the lattice spacings to each other. 
The MILC error is dominated 
by their continuum extrapolation. Our Figs~\ref{fig:fps} 
and~\ref{fig:a2dep} show very little dependence on lattice 
spacing (when using $w_0$) and benign extrapolations. 
Our error is nevertheless also dominated by the 
continuum extrapolation uncertainties along with statistical errors. 

Our analysis also gives results for the properties of 
the $\eta_s$ meson, by fixing its mass and decay constant 
in the continuum and physical light quark mass limits from those of the 
$\pi$ and $K$. Here the surprising result, found 
earlier in~\cite{Davies:2009tsa}, is how closely the properties 
of the $\eta_s$ match those expected from low order 
chiral perturbation theory. Our results here agree well 
with earlier results from $n_f=2+1$~\cite{Davies:2009tsa} 
as well as from earlier analysis on these $n_f=2+1+1$ 
configurations~\cite{Dowdall:2011wh}. 

We used $w_0/a$ to determine the relative lattice 
spacing between ensembles. Fixing the lattice spacing 
finally from $f_{\pi}$ gives a physical value for 
$w_0$ in Eq.~(\ref{eq:results}) of 0.1715(9) fm. 
This agrees at $2\sigma$ with the 
earlier result from BMW-c~\cite{Borsanyi:2012zs} 
of 0.1755(18) fm using the mass of the $\Omega$ baryon to 
fix the physical value. The error in the BMW-c result 
is dominated by the statistical errors in the lattice 
calculation of $M_{\Omega}$, and so we are able to 
obtain a smaller error using $f_{\pi}$.
  
In separate fits for comparison, not used 
for our central values, we also obtained 
values for $\sqrt{t_0}$ and $r_1$ in Eq.~(\ref{eq:results2}). 
Our $\sqrt{t_0}$ agrees with the value 
in~\cite{Borsanyi:2012zs}. The value for $r_1$ is not 
in good agreement with our earlier result, which 
it supersedes, on a subset 
of these ensembles~\cite{Dowdall:2011wh}, 
however. The reason for this is largely because the 
values for $r_1/a$ have been updated, resulting in 
changes outside the original error bars. This underscores the 
difficulty of determining parameters from the heavy quark 
potential accurately~\cite{Borsanyi:2012zs} and provides further incentive 
to use $w_0$.  

\section{Conclusions}
\label{sec:conclude}

We give here the most accurate result to date for 
$f_{K^+}/f_{\pi^+}$ from lattice QCD. 
Our result comes from `second-generation' gluon field 
configurations with a highly improved discretisation 
of QCD and including $u$, $d$, $s$, and $c$ quarks 
in the sea. We fit results from a range of ensembles 
with $u/d$ quark masses down to the physical point 
and including additional accurate information on the relative 
lattice spacing between the ensembles. We test the 
robustness of our continuum extrapolation using two different 
methods for lattice spacing determination. These features 
mean that we are able to improve on the error obtained 
by the MILC collaboration~\cite{Bazavov:2013cp} which aimed 
for a self-contained analysis using only physical $u/d$ quark 
masses. 

Our result is 
\begin{equation}
\frac{f_{K^+}}{f_{\pi^+}} = 1.1916(21).
\label{eq:finalres}
\end{equation}  
With this level of accuracy the difference between 
$f_K$ with $m_u=m_d$ and $f_{K^+}$, which we determine 
to be 0.27(7)\%,  is important. Going forward it will be necessary 
to make sure that $f_K$ and $f_{K^+}$ results from lattice 
QCD are averaged separately.  

Using Eq.~(\ref{eq:rat}) we determine 
\begin{equation}
\frac{|V_{us}|}{|V_{ud}|} = 0.23160(29)_{\mathrm{Br}(K^+)}(21)_{\mathrm{EM}}(41)_{\mathrm{latt}}.
\label{eq:vrat}
\end{equation} 
It is no longer true that the lattice QCD error is much larger 
than the total of experiment plus corrections to experiment 
from electromagnetism. 

Given a value of $V_{ud}$ from nuclear $\beta$ decay of 
0.97425(22)~\cite{Hardy:2008gy} gives 
\begin{equation}
|V_{us}| = 0.22564(28)_{\mathrm{Br}(K^+)}(20)_{\mathrm{EM}}(40)_{\mathrm{latt}}(5)_{V_{ud}}.
\end{equation}
This agrees well with values from experimental results for semileptonic $K$ 
decay rates combined with lattice QCD calculations of the appropriate 
hadronic form factor~\cite{Bazavov:2012cd, Boyle:2010bh, flag}.
The test of unitarity of the first row of the CKM matrix 
yields $1-|V_{ud}|^2-|V_{us}|^2-|V_{ub}|^2 = -0.00009(51)$. This 
agrees well with the Standard Model result of zero and 
pushes the scale of new physics 
above 10 TeV~\cite{Cirigliano:2009wk}. 
To improve the limit on this scale significantly now needs further improvements 
to the accuracy of $V_{ud}$~\cite{Hardy:2008gy}. 

\vspace{1mm}
\noindent{{\bf Acknowledgements}} We are grateful to the MILC collaboration for the use of their 
gauge configurations and to R. van de Water for useful discussions. 
We have used the MILC code for some of our propagator 
calculations.  
The results described here were obtained using the Darwin Supercomputer 
of the University of Cambridge High Performance 
Computing Service as part of STFC's DiRAC facility. 
We are grateful to the Darwin support staff for assistance. 
This work was funded by DFG (SFB-TR 55), NSF, the Royal Society, the Wolfson Foundation and STFC. 

\bibliography{fkpi}

\begin{thebibliography}{36}
\expandafter\ifx\csname natexlab\endcsname\relax\def\natexlab#1{#1}\fi
\expandafter\ifx\csname bibnamefont\endcsname\relax
  \def\bibnamefont#1{#1}\fi
\expandafter\ifx\csname bibfnamefont\endcsname\relax
  \def\bibfnamefont#1{#1}\fi
\expandafter\ifx\csname citenamefont\endcsname\relax
  \def\citenamefont#1{#1}\fi
\expandafter\ifx\csname url\endcsname\relax
  \def\url#1{\texttt{#1}}\fi
\expandafter\ifx\csname urlprefix\endcsname\relax\def\urlprefix{URL }\fi
\providecommand{\bibinfo}[2]{#2}
\providecommand{\eprint}[2][]{\url{#2}}

\bibitem[{\citenamefont{Beringer et~al.}(2012)}]{pdg}
\bibinfo{author}{\bibfnamefont{J.}~\bibnamefont{Beringer}} \bibnamefont{et~al.}
  (\bibinfo{collaboration}{Particle Data Group}), \bibinfo{journal}{Phys. Rev.}
  \textbf{\bibinfo{volume}{D86}}, \bibinfo{pages}{010001}
  (\bibinfo{year}{2012}).

\bibitem[{\citenamefont{Antonelli et~al.}(2010)\citenamefont{Antonelli,
  Cirigliano, Isidori, Mescia, Moulson et~al.}}]{flag}
\bibinfo{author}{\bibfnamefont{M.}~\bibnamefont{Antonelli}},
  \bibinfo{author}{\bibfnamefont{V.}~\bibnamefont{Cirigliano}},
  \bibinfo{author}{\bibfnamefont{G.}~\bibnamefont{Isidori}},
  \bibinfo{author}{\bibfnamefont{F.}~\bibnamefont{Mescia}},
  \bibinfo{author}{\bibfnamefont{M.}~\bibnamefont{Moulson}},
  \bibnamefont{et~al.}, \bibinfo{journal}{Eur.Phys.J.}
  \textbf{\bibinfo{volume}{C69}}, \bibinfo{pages}{399} (\bibinfo{year}{2010}),
  \eprint{1005.2323}.

\bibitem[{\citenamefont{Cirigliano and Neufeld}(2011)}]{Cirigliano:2011tm}
\bibinfo{author}{\bibfnamefont{V.}~\bibnamefont{Cirigliano}} \bibnamefont{and}
  \bibinfo{author}{\bibfnamefont{H.}~\bibnamefont{Neufeld}},
  \bibinfo{journal}{Phys.Lett.} \textbf{\bibinfo{volume}{B700}},
  \bibinfo{pages}{7} (\bibinfo{year}{2011}), \eprint{1102.0563}.

\bibitem[{\citenamefont{Marciano}(2004)}]{Marciano:2004uf}
\bibinfo{author}{\bibfnamefont{W.~J.} \bibnamefont{Marciano}},
  \bibinfo{journal}{Phys.Rev.Lett.} \textbf{\bibinfo{volume}{93}},
  \bibinfo{pages}{231803} (\bibinfo{year}{2004}), \eprint{hep-ph/0402299}.

\bibitem[{\citenamefont{Follana et~al.}(2008)\citenamefont{Follana, Davies,
  Lepage, and Shigemitsu}}]{Follana:2007uv}
\bibinfo{author}{\bibfnamefont{E.}~\bibnamefont{Follana}},
  \bibinfo{author}{\bibfnamefont{C.}~\bibnamefont{Davies}},
  \bibinfo{author}{\bibfnamefont{G.}~\bibnamefont{Lepage}}, \bibnamefont{and}
  \bibinfo{author}{\bibfnamefont{J.}~\bibnamefont{Shigemitsu}}
  (\bibinfo{collaboration}{HPQCD Collaboration, UKQCD Collaboration}),
  \bibinfo{journal}{Phys.Rev.Lett.} \textbf{\bibinfo{volume}{100}},
  \bibinfo{pages}{062002} (\bibinfo{year}{2008}), \eprint{0706.1726}.

\bibitem[{\citenamefont{Durr et~al.}(2010)\citenamefont{Durr, Fodor, Hoelbling,
  Katz, Krieg et~al.}}]{Durr:2010hr}
\bibinfo{author}{\bibfnamefont{S.}~\bibnamefont{Durr}},
  \bibinfo{author}{\bibfnamefont{Z.}~\bibnamefont{Fodor}},
  \bibinfo{author}{\bibfnamefont{C.}~\bibnamefont{Hoelbling}},
  \bibinfo{author}{\bibfnamefont{S.}~\bibnamefont{Katz}},
  \bibinfo{author}{\bibfnamefont{S.}~\bibnamefont{Krieg}},
  \bibnamefont{et~al.}, \bibinfo{journal}{Phys.Rev.}
  \textbf{\bibinfo{volume}{D81}}, \bibinfo{pages}{054507}
  (\bibinfo{year}{2010}), \eprint{1001.4692}.

\bibitem[{\citenamefont{Bazavov et~al.}(2010{\natexlab{a}})}]{Bazavov:2010hj}
\bibinfo{author}{\bibfnamefont{A.}~\bibnamefont{Bazavov}} \bibnamefont{et~al.}
  (\bibinfo{collaboration}{MILC Collaboration}), \bibinfo{journal}{PoS}
  \textbf{\bibinfo{volume}{LATTICE2010}}, \bibinfo{pages}{074}
  (\bibinfo{year}{2010}{\natexlab{a}}), \eprint{1012.0868}.

\bibitem[{\citenamefont{Aoki et~al.}(2011)}]{Aoki:2010dy}
\bibinfo{author}{\bibfnamefont{Y.}~\bibnamefont{Aoki}} \bibnamefont{et~al.}
  (\bibinfo{collaboration}{RBC Collaboration, UKQCD Collaboration}),
  \bibinfo{journal}{Phys.Rev.} \textbf{\bibinfo{volume}{D83}},
  \bibinfo{pages}{074508} (\bibinfo{year}{2011}), \eprint{1011.0892}.

\bibitem[{\citenamefont{Laiho and Van~de Water}(2011)}]{Laiho:2011np}
\bibinfo{author}{\bibfnamefont{J.}~\bibnamefont{Laiho}} \bibnamefont{and}
  \bibinfo{author}{\bibfnamefont{R.~S.} \bibnamefont{Van~de Water}},
  \bibinfo{journal}{PoS} \textbf{\bibinfo{volume}{LATTICE2011}},
  \bibinfo{pages}{293} (\bibinfo{year}{2011}), \eprint{1112.4861}.

\bibitem[{\citenamefont{Bazavov
  et~al.}(2013{\natexlab{a}})\citenamefont{Bazavov, Bernard, DeTar, Foley,
  Freeman et~al.}}]{Bazavov:2013cp}
\bibinfo{author}{\bibfnamefont{A.}~\bibnamefont{Bazavov}},
  \bibinfo{author}{\bibfnamefont{C.}~\bibnamefont{Bernard}},
  \bibinfo{author}{\bibfnamefont{C.}~\bibnamefont{DeTar}},
  \bibinfo{author}{\bibfnamefont{J.}~\bibnamefont{Foley}},
  \bibinfo{author}{\bibfnamefont{W.}~\bibnamefont{Freeman}},
  \bibnamefont{et~al.} (\bibinfo{collaboration}{MILC Collaboration}),
  \bibinfo{journal}{Phys.Rev. Lett.} \textbf{\bibinfo{volume}{110}},
  \bibinfo{pages}{172003} (\bibinfo{year}{2013}{\natexlab{a}}),
  \eprint{1301.5855}.

\bibitem[{\citenamefont{Follana et~al.}(2007)}]{Follana:2006rc}
\bibinfo{author}{\bibfnamefont{E.}~\bibnamefont{Follana}} \bibnamefont{et~al.}
  (\bibinfo{collaboration}{HPQCD collaboration}), \bibinfo{journal}{Phys.Rev.}
  \textbf{\bibinfo{volume}{D75}}, \bibinfo{pages}{054502}
  (\bibinfo{year}{2007}), \eprint{hep-lat/0610092}.

\bibitem[{\citenamefont{Hart et~al.}(2009)\citenamefont{Hart, {von Hippel}, and
  Horgan}}]{Hart:2008sq}
\bibinfo{author}{\bibfnamefont{A.}~\bibnamefont{Hart}},
  \bibinfo{author}{\bibfnamefont{G.~M.} \bibnamefont{{von Hippel}}},
  \bibnamefont{and} \bibinfo{author}{\bibfnamefont{R.~R.} \bibnamefont{Horgan}}
  (\bibinfo{collaboration}{HPQCD collaboration}), \bibinfo{journal}{Phys. Rev.}
  \textbf{\bibinfo{volume}{D79}}, \bibinfo{pages}{074008}
  (\bibinfo{year}{2009}), \eprint{0812.0503}.

\bibitem[{\citenamefont{Borsanyi et~al.}(2012)\citenamefont{Borsanyi, Durr,
  Fodor, Hoelbling, Katz et~al.}}]{Borsanyi:2012zs}
\bibinfo{author}{\bibfnamefont{S.}~\bibnamefont{Borsanyi}},
  \bibinfo{author}{\bibfnamefont{S.}~\bibnamefont{Durr}},
  \bibinfo{author}{\bibfnamefont{Z.}~\bibnamefont{Fodor}},
  \bibinfo{author}{\bibfnamefont{C.}~\bibnamefont{Hoelbling}},
  \bibinfo{author}{\bibfnamefont{S.~D.} \bibnamefont{Katz}},
  \bibnamefont{et~al.}, \bibinfo{journal}{JHEP}
  \textbf{\bibinfo{volume}{1209}}, \bibinfo{pages}{010} (\bibinfo{year}{2012}),
  \eprint{1203.4469}.

\bibitem[{\citenamefont{Bazavov et~al.}(2010{\natexlab{b}})}]{Bazavov:2010ru}
\bibinfo{author}{\bibfnamefont{A.}~\bibnamefont{Bazavov}} \bibnamefont{et~al.}
  (\bibinfo{collaboration}{MILC collaboration}), \bibinfo{journal}{Phys.Rev.}
  \textbf{\bibinfo{volume}{D82}}, \bibinfo{pages}{074501}
  (\bibinfo{year}{2010}{\natexlab{b}}), \eprint{1004.0342}.

\bibitem[{\citenamefont{Bazavov et~al.}(2013{\natexlab{b}})}]{Bazavov:2012uw}
\bibinfo{author}{\bibfnamefont{A.}~\bibnamefont{Bazavov}} \bibnamefont{et~al.}
  (\bibinfo{collaboration}{MILC Collaboration}), \bibinfo{journal}{Phys.Rev.}
  \textbf{\bibinfo{volume}{D87}}, \bibinfo{pages}{054505}
  (\bibinfo{year}{2013}{\natexlab{b}}), \eprint{1212.4768}.

\bibitem[{\citenamefont{Luscher}(2010)}]{Luscher:2010iy}
\bibinfo{author}{\bibfnamefont{M.}~\bibnamefont{Luscher}},
  \bibinfo{journal}{JHEP} \textbf{\bibinfo{volume}{1008}}, \bibinfo{pages}{071}
  (\bibinfo{year}{2010}), \eprint{1006.4518}.

\bibitem[{\citenamefont{Bazavov et~al.}(2010{\natexlab{c}})}]{Bazavov:2009bb}
\bibinfo{author}{\bibfnamefont{A.}~\bibnamefont{Bazavov}} \bibnamefont{et~al.},
  \bibinfo{journal}{Rev. Mod. Phys.} \textbf{\bibinfo{volume}{82}},
  \bibinfo{pages}{1349} (\bibinfo{year}{2010}{\natexlab{c}}),
  \eprint{0903.3598}.

\bibitem[{\citenamefont{Dowdall et~al.}(2012)}]{Dowdall:2011wh}
\bibinfo{author}{\bibfnamefont{R.}~\bibnamefont{Dowdall}} \bibnamefont{et~al.}
  (\bibinfo{collaboration}{HPQCD Collaboration}), \bibinfo{journal}{Phys.Rev.}
  \textbf{\bibinfo{volume}{D85}}, \bibinfo{pages}{054509}
  (\bibinfo{year}{2012}), \eprint{1110.6887}.

\bibitem[{\citenamefont{Davies et~al.}(2010)\citenamefont{Davies, Follana,
  Kendall, Lepage, and McNeile}}]{Davies:2009tsa}
\bibinfo{author}{\bibfnamefont{C.}~\bibnamefont{Davies}},
  \bibinfo{author}{\bibfnamefont{E.}~\bibnamefont{Follana}},
  \bibinfo{author}{\bibfnamefont{I.}~\bibnamefont{Kendall}},
  \bibinfo{author}{\bibfnamefont{G.}~\bibnamefont{Lepage}}, \bibnamefont{and}
  \bibinfo{author}{\bibfnamefont{C.}~\bibnamefont{McNeile}}
  (\bibinfo{collaboration}{HPQCD collaboration}), \bibinfo{journal}{Phys.Rev.}
  \textbf{\bibinfo{volume}{D81}}, \bibinfo{pages}{034506}
  (\bibinfo{year}{2010}), \eprint{0910.1229}.

\bibitem[{\citenamefont{Lepage et~al.}(2002)}]{gplbayes}
\bibinfo{author}{\bibfnamefont{G.~P.} \bibnamefont{Lepage}}
  \bibnamefont{et~al.}, \bibinfo{journal}{Nucl. Phys. Proc. Suppl.}
  \textbf{\bibinfo{volume}{106}}, \bibinfo{pages}{12} (\bibinfo{year}{2002}),
  \eprint{hep-lat/0110175}.

\bibitem[{\citenamefont{Lepage}()}]{fitcode}
\bibinfo{author}{\bibfnamefont{G.~P.} \bibnamefont{Lepage}},
  \emph{\bibinfo{title}{{https://github.com/gplepage}}}.

\bibitem[{\citenamefont{Sommer}(1994)}]{Sommer:1993ce}
\bibinfo{author}{\bibfnamefont{R.}~\bibnamefont{Sommer}},
  \bibinfo{journal}{Nucl. Phys.} \textbf{\bibinfo{volume}{B411}},
  \bibinfo{pages}{839} (\bibinfo{year}{1994}), \eprint{hep-lat/9310022}.

\bibitem[{\citenamefont{McNeile et~al.}(2013)\citenamefont{McNeile, Bazavov,
  Davies, Dowdall, Hornbostel et~al.}}]{McNeile:2012xh}
\bibinfo{author}{\bibfnamefont{C.}~\bibnamefont{McNeile}},
  \bibinfo{author}{\bibfnamefont{A.}~\bibnamefont{Bazavov}},
  \bibinfo{author}{\bibfnamefont{C.}~\bibnamefont{Davies}},
  \bibinfo{author}{\bibfnamefont{R.}~\bibnamefont{Dowdall}},
  \bibinfo{author}{\bibfnamefont{K.}~\bibnamefont{Hornbostel}},
  \bibnamefont{et~al.}, \bibinfo{journal}{Phys.Rev.}
  \textbf{\bibinfo{volume}{D87}}, \bibinfo{pages}{034503}
  (\bibinfo{year}{2013}), \eprint{1211.6577}.

\bibitem[{\citenamefont{Sharpe and Shoresh}(2000)}]{Sharpe:2000bc}
\bibinfo{author}{\bibfnamefont{S.~R.} \bibnamefont{Sharpe}} \bibnamefont{and}
  \bibinfo{author}{\bibfnamefont{N.}~\bibnamefont{Shoresh}},
  \bibinfo{journal}{Phys.Rev.} \textbf{\bibinfo{volume}{D62}},
  \bibinfo{pages}{094503} (\bibinfo{year}{2000}), \eprint{hep-lat/0006017}.

\bibitem[{\citenamefont{Colangelo et~al.}(2005)\citenamefont{Colangelo, Durr,
  and Haefeli}}]{Colangelo:2005gd}
\bibinfo{author}{\bibfnamefont{G.}~\bibnamefont{Colangelo}},
  \bibinfo{author}{\bibfnamefont{S.}~\bibnamefont{Durr}}, \bibnamefont{and}
  \bibinfo{author}{\bibfnamefont{C.}~\bibnamefont{Haefeli}},
  \bibinfo{journal}{Nucl.Phys.} \textbf{\bibinfo{volume}{B721}},
  \bibinfo{pages}{136} (\bibinfo{year}{2005}), \eprint{hep-lat/0503014}.

\bibitem[{\citenamefont{Aubin and Bernard}(2003)}]{Aubin:2003uc}
\bibinfo{author}{\bibfnamefont{C.}~\bibnamefont{Aubin}} \bibnamefont{and}
  \bibinfo{author}{\bibfnamefont{C.}~\bibnamefont{Bernard}},
  \bibinfo{journal}{Phys.Rev.} \textbf{\bibinfo{volume}{D68}},
  \bibinfo{pages}{074011} (\bibinfo{year}{2003}), \eprint{hep-lat/0306026}.

\bibitem[{\citenamefont{Amoros et~al.}(2001)\citenamefont{Amoros, Bijnens, and
  Talavera}}]{Amoros:2001cp}
\bibinfo{author}{\bibfnamefont{G.}~\bibnamefont{Amoros}},
  \bibinfo{author}{\bibfnamefont{J.}~\bibnamefont{Bijnens}}, \bibnamefont{and}
  \bibinfo{author}{\bibfnamefont{P.}~\bibnamefont{Talavera}},
  \bibinfo{journal}{Nucl.Phys.} \textbf{\bibinfo{volume}{B602}},
  \bibinfo{pages}{87} (\bibinfo{year}{2001}), \eprint{hep-ph/0101127}.

\bibitem[{\citenamefont{Basak et~al.}(2012)}]{Basak:2012zx}
\bibinfo{author}{\bibfnamefont{S.}~\bibnamefont{Basak}} \bibnamefont{et~al.}
  (\bibinfo{collaboration}{MILC Collaboration}), \bibinfo{journal}{PoS}
  \textbf{\bibinfo{volume}{LATTICE2012}}, \bibinfo{pages}{137}
  (\bibinfo{year}{2012}), \eprint{1210.8157}.

\bibitem[{\citenamefont{Portelli et~al.}(2011)\citenamefont{Portelli, Durr,
  Fodor, Frison, Hoelbling et~al.}}]{Portelli:2012pn}
\bibinfo{author}{\bibfnamefont{A.}~\bibnamefont{Portelli}},
  \bibinfo{author}{\bibfnamefont{S.}~\bibnamefont{Durr}},
  \bibinfo{author}{\bibfnamefont{Z.}~\bibnamefont{Fodor}},
  \bibinfo{author}{\bibfnamefont{J.}~\bibnamefont{Frison}},
  \bibinfo{author}{\bibfnamefont{C.}~\bibnamefont{Hoelbling}},
  \bibnamefont{et~al.}, \bibinfo{journal}{PoS}
  \textbf{\bibinfo{volume}{LATTICE2011}}, \bibinfo{pages}{136}
  (\bibinfo{year}{2011}), \eprint{1201.2787}.

\bibitem[{\citenamefont{Blum et~al.}(2010)\citenamefont{Blum, Zhou, Doi,
  Hayakawa, Izubuchi et~al.}}]{Blum:2010ym}
\bibinfo{author}{\bibfnamefont{T.}~\bibnamefont{Blum}},
  \bibinfo{author}{\bibfnamefont{R.}~\bibnamefont{Zhou}},
  \bibinfo{author}{\bibfnamefont{T.}~\bibnamefont{Doi}},
  \bibinfo{author}{\bibfnamefont{M.}~\bibnamefont{Hayakawa}},
  \bibinfo{author}{\bibfnamefont{T.}~\bibnamefont{Izubuchi}},
  \bibnamefont{et~al.}, \bibinfo{journal}{Phys.Rev.}
  \textbf{\bibinfo{volume}{D82}}, \bibinfo{pages}{094508}
  (\bibinfo{year}{2010}), \eprint{1006.1311}.

\bibitem[{\citenamefont{Bazavov et~al.}(2009{\natexlab{a}})}]{Bazavov:2009tw}
\bibinfo{author}{\bibfnamefont{A.}~\bibnamefont{Bazavov}} \bibnamefont{et~al.}
  (\bibinfo{collaboration}{MILC Collaboration}), \bibinfo{journal}{PoS}
  \textbf{\bibinfo{volume}{LAT2009}}, \bibinfo{pages}{079}
  (\bibinfo{year}{2009}{\natexlab{a}}), \eprint{0910.3618}.

\bibitem[{\citenamefont{Bazavov et~al.}(2009{\natexlab{b}})}]{Bazavov:2009ir}
\bibinfo{author}{\bibfnamefont{A.}~\bibnamefont{Bazavov}} \bibnamefont{et~al.}
  (\bibinfo{collaboration}{MILC Collaboration}), \bibinfo{journal}{PoS}
  \textbf{\bibinfo{volume}{LAT2009}}, \bibinfo{pages}{077}
  (\bibinfo{year}{2009}{\natexlab{b}}), \eprint{0911.0472}.

\bibitem[{\citenamefont{Hardy and Towner}(2009)}]{Hardy:2008gy}
\bibinfo{author}{\bibfnamefont{J.}~\bibnamefont{Hardy}} \bibnamefont{and}
  \bibinfo{author}{\bibfnamefont{I.}~\bibnamefont{Towner}},
  \bibinfo{journal}{Phys.Rev.} \textbf{\bibinfo{volume}{C79}},
  \bibinfo{pages}{055502} (\bibinfo{year}{2009}), \eprint{0812.1202}.

\bibitem[{\citenamefont{Bazavov
  et~al.}(2013{\natexlab{c}})\citenamefont{Bazavov, Bernard, Bouchard, DeTar,
  Du et~al.}}]{Bazavov:2012cd}
\bibinfo{author}{\bibfnamefont{A.}~\bibnamefont{Bazavov}},
  \bibinfo{author}{\bibfnamefont{C.}~\bibnamefont{Bernard}},
  \bibinfo{author}{\bibfnamefont{C.}~\bibnamefont{Bouchard}},
  \bibinfo{author}{\bibfnamefont{C.}~\bibnamefont{DeTar}},
  \bibinfo{author}{\bibfnamefont{D.}~\bibnamefont{Du}}, \bibnamefont{et~al.}
  (\bibinfo{collaboration}{MILC Collaboration}), \bibinfo{journal}{Phys.Rev.}
  \textbf{\bibinfo{volume}{D87}}, \bibinfo{pages}{073012}
  (\bibinfo{year}{2013}{\natexlab{c}}), \eprint{1212.4993}.

\bibitem[{\citenamefont{Boyle et~al.}(2010)}]{Boyle:2010bh}
\bibinfo{author}{\bibfnamefont{P.}~\bibnamefont{Boyle}} \bibnamefont{et~al.}
  (\bibinfo{collaboration}{RBC-UKQCD Collaboration}),
  \bibinfo{journal}{Eur.Phys.J.} \textbf{\bibinfo{volume}{C69}},
  \bibinfo{pages}{159} (\bibinfo{year}{2010}), \eprint{1004.0886}.

\bibitem[{\citenamefont{Cirigliano et~al.}(2010)\citenamefont{Cirigliano,
  Jenkins, and Gonzalez-Alonso}}]{Cirigliano:2009wk}
\bibinfo{author}{\bibfnamefont{V.}~\bibnamefont{Cirigliano}},
  \bibinfo{author}{\bibfnamefont{J.}~\bibnamefont{Jenkins}}, \bibnamefont{and}
  \bibinfo{author}{\bibfnamefont{M.}~\bibnamefont{Gonzalez-Alonso}},
  \bibinfo{journal}{Nucl.Phys.} \textbf{\bibinfo{volume}{B830}},
  \bibinfo{pages}{95} (\bibinfo{year}{2010}), \eprint{0908.1754}.

\end{thebibliography}

\end{document}